\documentclass{ws-p8-50x6-00}
\def\rightnote{{\it \ \ Quantum Gravity at the GR16}}
\begin{document}
\title{GR16: Quantum General Relativity}
\author{Carlo Rovelli}

\address{Physics Department, University of Pittsburgh, PA-15260, USA\\
Centre de Physique Theorique, Luminy, Marseille, F-13288, EU\\ 
E-mail: rovelli@cpt.univ-mrs.fr}

\maketitle \abstracts{This is the report of the 
{\em Quantum General Relativity\/} session, at the 16th International
Conference on General Relativity \& Gravitation, held on July 15th to
21st 2001, in Durban, South Africa.  This report will appear on the
Proceedings of the conference.}

\section{Highlights}

Current research in quantum gravity is very lively.  The research is
actively developing both towards the construction of the fundamental
theory and towards applications.

A large part of the efforts to develop the general theory of the
quantum gravitational field is presently concentrated in {\em loop
quantum gravity,} in its various forms.  The general theory of loop
quantum gravity was reviewed at this conference in a plenary talk by
Abhay Ashtekar.  In the session, several specific aspects of the
theory were discussed.  The current efforts to understand the low
energy behavior of loop quantum gravity were reviewed by Abhay
Ashtekar and Jerzy Lewandowski.  The covariant spacetime version of
the theory, denoted {\em spinfoam\/} theory, and some notable
finiteness results due also to Alejandro Perez and Louis Crane were
illustrated by myself.  Various aspects of the discreteness of the
geometry predicted by loop quantum gravity were discussed by Seth
Major.  Unfortunately, Martin Bojowald had to cancel his participation
to the conference at the last minute, and his scheduled communication
on his exciting recent results on the application of loop quantum
gravity to the initial singularity had to be cancelled.  The
simplicial approach, whose recent developments were reviewed by Ruth
Williams, has also recently converged with loop quantum gravity.  In
fact, on the one hand the spinfoam formalism include the Barrett Crane
simplicial model, on the other hand it can be viewed as the covariant
version of (canonical) loop quantum gravity.  Altogether, loop
quantum gravity appears to be in a very exciting phase of intense
development and novel results.

Other approaches to quantum gravity were considered in the session as
well.  In particular: non commutative geometry (Kamesh Wali), the
string motivated AdS/CFT theory (Robert Mann), lattice and functional
formalisms (Renate Loll).

The most studied applications of quantum gravity are in black hole
physics, gravitational collapse, and cosmological models.  The
application of quantum gravity to gravitational collapse is a long
studied problem.  A new and rather surprising results presented by
Peter H\'aj{\'\i}\v cek, obtained via a careful hamiltonian analysis
of the problem came therefore as a surprise: a shell bounces and
re-expands, and the singularity disappears.  The collapse problem was
discussed also by Mike Ryan, Cenalo Vaz and Valentin Gladush.

The most characteristic physical system for quantum gravity is
certainly the quantum black hole.  There are now two well-known
derivations of the Bekenstein-Hawking black hole entropy from a
complete quantum gravity theory, obtained respectively within string
theory and within loop quantum gravity.  But a complete physical
understanding of the quantum black hole thermal behavior is not yet
available, and the problem keeps raising interest.  The problem was
discussed in the session by Steven Carlip, Gabor Kunstatter and Louis
Witten.  Finally, a cosmological model with a perfect fluid was discussed by
Julio Fabris.  

I'll now illustrate all the contributions to the session in more
detail.

\section{General theory: loop quantum gravity}

\begin{description}

\item[\em Abhay Ashtekar: ``Semiclassical issues in nonperturbative QG" ] 
Ashtekar discussed the use of a detailed mathematical theory of
quantum geometry, in order to analyze the fundamental discreteness of
spacetime predicted by loop quantum gravity.  This discreteness makes
it difficult to relate the underlying `polymer excitations' of quantum
geometry to the Fock states normally used in low energy physics.  To
bridge this gap, using some recent results of Varadarajan, Ashtekar
and Lewandowski have located, analyzed and used Minkowskian Fock
states in the background independent framework.  The fundamental
discreteness is lost in the standard semi-classical description
because Fock states constitute only a small subset of all
non-perturbative states.  While the quantum geometry operators have
purely discrete eigenvalues in the full theory, none of these
eigenvectors belong to the Fock space.  This framework is well-suited
to address two key questions: Can the background-independent,
non-perturbative theory reproduce the familiar low-energy physics on,
say, suitable coarse graining?  and, Can one pin-point where and why
perturbation theory fails?  Ashtekar described work in progress on
both these fronts. 

\item[\em Carlo Rovelli: ``Spin foam models" ]  A spin foam model can be
seen as the covariant, or Feynman integral, version of a loop quantum
gravity theory.  A spin foam represents a four-geometry and the model
is formulated as a sum over these four-geometries.  Remarkably, there
is a duality between these models and certain particular field
theories defined on group manifolds (Group Field Theory, or GFT).  The
sum over spin foam can be generated as the Feynman perturbative
expansion of the GFT. Each spacetime appears therefore as the Feynman
graph of the auxiliary GFT theory.  This phenomenon is a 4d analog of
a phenomenon first noticed in 2d quantum gravity, or ```zero
dimensional string theory", where the spacetime could be obtained as
Feynman graphs of an auxiliary matrix model.  The main result
presented is that (up to certain degenerate terms) the theory is {\em
finite\/} order by order in the perturbation expansion.  This allows
finite three-geometry to three-geometry transition amplitudes to be
perturbatively computed explicitly in the physical, lorentzian, 4d
theory.

\item[\em Seth Major: ``Discrete geometry" ]  Major presented recent
results on the spectra of the angle, area, and volume operators,
including limits placed on the level of discreteness by semiclassical
behavior and classic tests of fundamental physics.

\item[\em Ruth Williams: ``Progress in discrete quantum gravity" ] 
Approaches to discrete quantum gravity include sums over histories
using the simplicial analogue of the Einstein action formulated by
Regge, and state sum models, such as the Ponzano-Regge and Turaev-Viro
models in three dimensions and the Barrett-Crane model in four
dimensions.  There has been a remarkable convergence with the loop
quantum gravity approach, in recent years.  Williams described
progress with simplicial and state sum models, including a discrete
version of the calculation of the ground state wave function for
linearized gravity, and an investigation of perturbations of
three-dimensional state sum models.

\end{description}

\section{General theory: others}
\begin{description}

\item[\em Kamesh Wali: ``Action functionals on a two-sheeted spacetime" ]
Wali proposes a model of left- and right- chiral fields living on the
two-sheeted space-time along with two distinct gauge fields, within
the framework of non-commutative geometry.  The mathematics presented
generalizes the fundamental concepts of Riemannian geometry, such as
differential forms, connection and curvature.  The model has a rich
and complex structure with new interaction terms to be explored.  One
of them is a parity violating interaction arising from the
gravitational sector of their generalized theory.

\item[\em Robert Mann: ``Conserved quantities in AdS/CFT" ]  The talk
reported on work concerning the problem of defining conserved charges
in gravitational theory.  This is a long-standing problem in general
relativity -- the equivalence principle implies that locally gravity
cannot be distinguished from acceleration, and the problem of defining
a local measure of gravitational energy and angular momentum has
remained elusive.  A reasonable measure of success has been attained
in recent years using the Brown/York quasilocal formalism, which
proposes a definition of conserved quantities enclosed by a $(d-2)$
surface in $d$ dimensions.  A generic problem with the quasilocal
formalism is that the conserved quantities typically diverge as the
mean radial size of the enclosed surface becomes infinite.  A recent
proposal inspired by the AdS/CFT correspondence offers a potential
resolution to this difficulty.  This conjectured correspondence states
that a quantum theory of gravity whose boundary conditions are the
same as those of asymptotically anti de Sitter (AdS) spacetime is in
1-1 correspondence with a quantum conformal field theory (CFT) defined
on the boundary of this spacetime.  Quantum field theories generically
have counterterms, and the correspondence suggests that these
counterterms should be geometric invariants of the induced metric on
the boundary at infinity.  Employing this proposal, Mann computed the
energy, angular momentum and entropy of spacetimes containing rotating
black holes with NUT charge (ie in the general Kerr-NUT-anti de Sitter
class).  He found in all cases that the conserved quantities were
finite.  No reference spacetime was required at any stage of the
calculation.  He then showed how to extend this prescription to
asymptotically flat spacetimes, and obtained corresponding results
that are finite.  The interpretation of these quantities in
spacetimes with NUT charge is still obscure.

\item[\em Renate Loll: ``The conformal factor problem" ] Under certain
assumption on the behavior of the partition function under
renormalization, one finds the interesting result that the well known
divergence due to the conformal modes is cancelled by a Faddeev-Popov
determinant.  This result confirms that absence of the conformal
sickness noticed by Loll in the lorentzian lattice models.

\end{description}

\section{Gravitational collapse}

\begin{description}
\item[\em Peter H\'aj{\'\i}\v cek: ``Gravitational collapse" ]
H\'aj{\'\i}\v cek studies the quantum dynamics of a self-gravitating
null shell for the spherically symmetric case, using only gauge
invariant quantities (Dirac observables).  The definition of evolution
is based on an asymptotically non-trivial symmetry of the model (time
translation element of the BMS group).  H\'aj{\'\i}\v cek proves in
particular, that a version of quantum theory exists, in which the
shell bounces and re-expands, and in which the singularity disappears. 
The quantum metric outside the shell contains a time-dependent mixture
of black and white holes.  These remarkable and surprising results
raised much interest and a lively discussion, in the session.

\item[\em Michael Ryan: ``Quantum collapse of dust shell: solutions" ] 
A hamiltonian formulation of the problem of the collapse of a dust
shell given by H\'aj{\'\i}\v cek and Kijowski and Kucha\v r allows one
to study the the minisuperspace problem of the quantum collapse of a
spherically symmetric dust shell, albeit with a very complicated
Hamiltonian operator.  Ryan presented both analytic approximate
solutions and exact numerical solutions to the problem.  Since the
solutions involve only geometrical variables defined on the shell, it
is necessary to reconstruct the quantum spacetime geometry around the
shell (possible in this special spherically symmetric case) in order
to discuss the interpretation of the solutions in terms of spacetime
geometry, especially the position of the shell with respect to
horizons.  This allowed Ryan to consider the quantum evolution of the
shell in spacetime.

\item[\em Cenalo Vaz: ``LeMaitre-Tolman-Bondi collapse" ]  Vaz applied the
Kuchar transformation to LeMaitre-Tolman-Bondi models of inhomogeneous
dust collapse.  Dirac's constraint quantization leads to a simple
Wheeler-DeWitt equation with a general mass function.  A solution was
presented over the entire class of models considered.  These solutions
cover collapse into both black holes and naked singularities (up to
the Cauchy Horizon) and can be employed in the study of quantum
gravitational effects in the final stages of collapse with different
initial conditions as well as, it is hoped, in the study of
inhomogeneous cosmologies.

\item[\em Valentin Gladush: ``Quasiclassical spherical
configuration" ]  Gladush has considered a quasi-classical model of
self-gravitating spherical dust shell, and studied the energy
spectrum using quasi-classical quantization rules, and the stability
of the system.

\end{description}

\section{Black holes}
\begin{description}
\item[\em Steven Carlip: ``Boundary conditions, constraints, BH entropy" ]  
Recent work by a number of authors has suggested that black hole
entropy may have a microscopic description in terms of a boundary
conformal field theory, living either near the horizon or at spatial
infinity.  Carlip described a new approach to boundary conditions near
such a boundary, in which the fall-off requirements are introduced as
a set of second class constraints that modify the algebra of
diffeomorphisms.  For Liouville theory in a dynamical background, this
approach leads to the standard central charge; for dilaton gravity,
the result is a Virasoro algebra with a classical central charge of
the form required to explain the Bekenstein-Hawking entropy.

\item[\em Gabor Kunstatter: ``Quantum Mechanics of Charged BHs" ]
Basic properties of black hole thermodynamics were used to quantize
generic charged black holes in the Euclidean sector.  The analysis led
to a discrete spectrum for the horizon area that is qualitatively
consistent with the results of Bekenstein, Mukhanov and others, but
different from the spectrum predicted by loop quantum gravity.  In
particular, the area spectrum was shown generically to be equally
spaced.  Quantization of the charge sector gave the condition $Q=me$,
for integer $m$, as expected, but a consistency condition constrained
the fine structure constant $e^2/\hbar$ to be a rational number.  Near
extremal black holes were shown to be highly quantum objects. 
Moreover, with a standard choice of factor ordering, extremal black
holes did not appear in the physical spectrum at all.

\item[\em Louis Witten: ``BH Quantization in Canonical Gravity" ] 
Witten started from the observation that the Schwarzschild black hole
can be viewed as a special case of the marginally bound
LeMaitre-Tolman-Bondi models of dust collapse corresponding to a
constant mass function.  Using a midi-superspace quantization of the
collapse for an arbitrary mass function illustrated in this same
session by C Vaz, Witten showed that the solution leads both to
Bekensteins's area spectrum for black holes as well as to the black
hole entropy.  In this context, the entropy is naturally interpreted
as the loss of information of the original matter distribution within
the collapsing cloud.

\end{description}

\section{Finite dimensional models}

\begin{description}

\item[\em Julio Fabris: ``Quantum cosmological perfect fluid" ]   Fabris
quantized the gravity-perfect fluid system, employing the Schutz's
formalism.  The matter degrees of freedom allow to recover the notion
of time.  Wave-functions are obtained in closed form, in the
mini-superspace, and the expectation value for the scale factor is
computed, yielding a singularity-free, bouncing Universe.  He showed
that these results are recovered by a classical system where, besides
the ordinary matter employed in the quantization procedure, a
universal repulsive fluid with a stiff matter equation of state is
added. 

\end{description}
\end{document}